\ifdefined\TWOCOLS
\documentclass[a4paper,twocolumn,12pt]{extarticle}
\else
\documentclass[a4paper,12pt]{extarticle}
\fi
\usepackage[utf8]{inputenc}
\usepackage{amsmath}
\usepackage{amsfonts}
\usepackage{amssymb}
\usepackage{graphicx}
\usepackage{caption}
\usepackage{subcaption}
\usepackage{epstopdf}
\usepackage{abstract}
\usepackage[toc,page]{appendix}
\usepackage[sort]{cite}
\usepackage{color}

\usepackage{algorithm}
\usepackage{algorithmicx}
\usepackage{algpseudocode}
\usepackage{setspace}
\usepackage{authblk}
\usepackage[round,numbers]{natbib}
\usepackage{authblk}

\usepackage{hyperref}

\newcommand{\half}[0]{\frac{1}{2}}
\newcommand{\bvec}[1]{\mathbf{#1}}

\newcommand{\dd}[0]{\ensuremath{\operatorname{d}}}
\newcommand{\eg}[0]{\emph{e.g.}}

\renewcommand{\cite}[1]{\citep{#1}}
\renewcommand{\thesubfigure}{\Alph{subfigure}}

\graphicspath{{./images/}}

\title{ {A method for molecular dynamics on curved surfaces} }
\author[1]{S. Paquay}
\author[1]{R. Kusters}
\affil[1]{Department of Applied Physics, Eindhoven University of Technology, Eindhoven, The Netherlands}
\date{  }

\begin{document}
\ifdefined\TWOCOLS
\twocolumn[
\begin{@twocolumnfalse}
\fi
  \maketitle
  
  \begin{abstract}
      \noindent Dynamics simulations of constrained particles can greatly aid in understanding the temporal and spatial evolution of biological processes such as lateral transport along membranes and self-assembly of {viruses}. Most theoretical efforts in the field of diffusive transport have focussed on solving the diffusion equation on curved surfaces, for which it is not tractable to incorporate particle interactions even though these play a crucial role in crowded systems. We show here that it is possible to combine standard constraint algorithms with the classical velocity Verlet scheme to perform molecular dynamics simulations of particles constrained to an arbitrarily curved surface, in which such interactions can be taken into account. Furthermore, unlike Brownian dynamics schemes in local coordinates, our method is based on Cartesian coordinates allowing for the reuse of many other standard tools without modifications, including parallelisation through domain decomposition. We show that by applying the schemes to the Langevin equation for various surfaces, confined Brownian motion is obtained, which has direct applications to many biological and physical problems. Finally we present two practical examples that highlight the applicability of the method: (i) the influence of crowding and shape on the lateral diffusion of proteins in curved membranes and (ii) the self-assembly of a coarse-grained virus capsid protein model.

%     More specifically, we show that the RATTLE constraint algorithm applied to single particle constraints is computationally very efficient and easy to implement.     
    \end{abstract}
\ifdefined\TWOCOLS
\end{@twocolumnfalse}
]
\fi

\setlength{\parindent}{1em}
\section{Introduction}

Diffusion is of paramount interest in the context of temporal and spatial evolution of biological systems. To name but one example, lateral diffusion along the plasma membrane \cite{jaskolski-2009}, a process ubiquitous in biological systems \cite{hofling-2013}, is crucial in the regulation of, \eg, synaptic strength regulation in neurons \cite{jackson-2011,straub-2012} and the regulation of the photosynthetic electron transport site in grana thylakoids \cite{tremmel-2003}. Many experimental \cite{ashby-2006,bloodgood-2005,hayashi-2005,vladimirou-2009} and theoretical \cite{tremmel-2003,holcman-20142,kusters-2013,kusters-2014} efforts have been made to understand how membrane shape and composition regulates protein diffusion on highly curved membrane structures. Most theoretical models are based on solving the diffusion equation on the curved surface \cite{benichou-2014, holcman-20142,kusters-2013,kusters-2014}. This method, however, is not always tractable, especially when complex particle-particle interactions, which are of great importance in these systems, are involved \cite{hofling-2013}.

Recently, a Brownian dynamics algorithm was developed aimed at describing the motion of mutually interacting particles on curved manifolds by Villareal \emph{et al.} \cite{villarreal-2014}. They briefly explored adding a restoring harmonic potential to pull particles towards the manifold but this idea was abandoned because a spring constant sufficiently high to constrain the particles would severely inhibit the allowed time step size. They opted instead for a solution in local coordinates.

We show here that instead of a harmonic ``spring'' to constrain the particles, one can use a standard  constraint algorithm \cite{ryckaert-1977,andersen-1983} to take into account the constraints of the manifold when solving the equations of motion without transforming to local coordinates. This has some advantages.
Firstly, the method works in Cartesian rather than local coordinates, and makes it possible to reuse many tools of the trade from molecular dynamics (MD) without any modifications, including standard Langevin approaches to model Brownian dynamics \cite{schneider-1978}.
Secondly, the constraint algorithm can be applied to only some of the particles in the system, allowing others to move freely throughout the volume{. This way,} interaction between particles diffusing on a manifold with those in the surrounding liquid can be simulated, which can be used to study how crowding and hydrodynamic effects around the cell membrane affect diffusion along the membrane \cite{ando-2010}.

An additional advantage of the method is that one can also study the self-assembly behaviour of coarse-grained molecules consisting of a few beads with one end constrained to a spherical template as a more complete model for self-assembly studies of certain viruses, \eg{}, immature HIV-1 \cite{goicochea-2011,hagan-2015}.
Although not particularly relevant for biological systems, the method allows incorporation of inertia effects should they be important. Note, however, that even if inertia is irrelevant, there are technical reasons to prefer Langevin dynamics over Brownian dynamics (the overdamped limit), as explained in section 1.8 of reference \cite{ladd-2009}.

In the remainder of this paper we describe RATTLE \cite{andersen-1983}, our constraint algorithm of choice, and present its specific implementation to the single-particle constraints of the curved surface in section \ref{sec:methods}. In section \ref{sec:results} we first verify how well RATTLE performs and if it conserves the total energy of the system. We then show that it can reproduce Brownian diffusion on manifolds, which we apply to determine escape times of particles in a crowded grana thylakoid model. As a second example we apply the method to study the self-assembly of a model virus capsid. Finally, in section \ref{sec:conclusion}, we succinctly present the most important conclusions from this paper and describe where the method can be applied.

\section{Methods}
\label{sec:methods}
In this section we present the equations of motion for particles constrained to curved surfaces. The numerical scheme is then obtained by applying a constraint algorithm in combination with the velocity Verlet algorithm to the derived equations.

\subsection{Equations of motion}
For unconstrained systems a MD simulation consists of solving Newton's equations of motion, which can be obtained from the Hamiltonian of the system. If the particles are all constrained to some arbitrary manifold, this can be incorporated into the Hamiltonian by means of introducing Lagrange multipliers $\lambda_i,$
\begin{align}
  \mathcal{H} =& \sum \limits_{i=1}^{N} \left[ \frac{1}{2 m_i}\bvec{p}_i^2 + \sum \limits_{j=i+1}^{N} V_{ij} + \lambda_i g(\bvec{x}_i) \right],\label{eqn:Hamiltonian-constrained}
\end{align}
where the function $g$ is chosen such that $g(\bvec{x}_i) = 0$ for all $i$ if the particles obey the constraint, with $\bvec{x}_i$ the position vector of particle $i,$ $\bvec{p}_i$ its momentum and $m_i$ its mass, and 
where the inter-particle potential $V_{ij}$ is a function of $\bvec{x}_i$ and $\bvec{x}_j,$ and $V_{ij} = V_{ji}.$ For instance, if we define $g(\bvec{x}_i) = \bvec{x}_i^2-R^2$ this constrains the particle positions $\bvec{x}_i$ to a sphere of radius $R.$

From Eq. \eqref{eqn:Hamiltonian-constrained} the equations of motion become
\begin{align}
  \frac{\dd \bvec{x}_i}{\dd t} = \frac{\partial \mathcal{H} }{\partial \bvec{p}_i} =& \frac{1}{m_i} \bvec{p}_i,  \label{eqn:Hamiltonian-q} \\
 \frac{\dd \bvec{p}_i}{\dd t} = -\frac{\partial \mathcal{H}}{\partial \bvec{x}_i} =& - \sum \limits_{j\neq i} \frac{\partial V_{ij}}{\partial \bvec{x}_i} - \lambda_i \frac{\partial g}{\partial \bvec{x}_i}, \label{eqn:Hamiltonian-p}
\end{align}
where $t$ is the time.
The expression for the change in position remains unchanged, but an additional term enters Eq. \eqref{eqn:Hamiltonian-p} for the change in momentum. Note also that $-\sum_{j\neq i}^N \partial V_{ij} / \partial \bvec{x}_i$ is the total force acting on particle $i,$ which we will denote as $\bvec{f}_i.$ Any force not generated by a potential can be added to this function, \eg{}, thermal fluctuations for Brownian dynamics. The term $\partial g/\partial \bvec{x}_i$ is simply the normal of the manifold, $\bvec{n}(\bvec{x}_i) := \bvec{n}_i,$ which allows Eq. \eqref{eqn:Hamiltonian-p} to be rewritten as
\begin{equation}
  \frac{ \dd \bvec{p}_i}{ \dd t } = \bvec{f}_i - \lambda_i \bvec{n}_i. \label{eqn:Hamiltonian-p-2}
\end{equation}

It is possible to derive a closed expression for the Lagrange multipliers in \eqref{eqn:Hamiltonian-p-2} as is done in, \eg, \cite{vest-2014}. However, the idea behind RATTLE is to iteratively determine them during a simulation. This makes the method more flexible, as it can also be applied for cases in which a closed expression for $\lambda$ is difficult to derive.

\subsection{Numerical scheme}
\label{sec:schemes}
We apply the standard RATTLE algorithm \cite{andersen-1983}{, explained in detail in \emph{e.g.} Ref. \cite{leimkuhler-boek},}  to Eqs. \eqref{eqn:Hamiltonian-q} and \eqref{eqn:Hamiltonian-p-2} for the specific case of a curved surface, enforcing the constraints $g(\bvec{x}_i)=0$ and $m_i^{-1} \bvec{p}_i\cdot\bvec{n}_i = \bvec{v}_i \cdot \bvec{n}_i=0,$ with the second constraint enforcing that the velocity component directed out of the surface should be 0.
{ This leads the pseudocode presented in Algorithm \ref{alg:rattle}.}
In the pseudocode, $\bvec{J}_i$ represents the Jacobi matrix of $\bvec{r}_i$ and $\bvec{I}$ is a $3 \times 3$ identity matrix. The other symbols are introduced below.

As mentioned before, RATTLE iteratively determines the Lagrange multipliers so that at the next time step, both constraints are satisfied and the new particle position and momentum are consistent with the total force acting upon it, including the constraint force $-\lambda_i \bvec{n}_i.$ This is done using Newton iteration. Let a superscript $m$ denote the current time step and superscript $m+1$ the next. Then Newton iteration constructs an approximation for both $\bvec{x}_{i}^{m+1}$ and $\lambda_i$ that simultaneously satisfy the equations $\bvec{x}_i^m - \bvec{x}_i^{m+1} + \Delta t \left( \bvec{p}_i^m + m_i^{-1} \Delta t(\bvec{f}_i^m - \lambda_i\bvec{n}_i^m)/2  \right) = \bvec{0}$ and $g(\bvec{x}_i^{m+1}) = 0.$ Because Newton iteration constructs a numerical approximation, it will never find the exact solution to these two equations. Instead, we iterate until the norm of the so-called residual vector $\bvec{r}_i:= \left(\bvec{r}_{i,x}, r_{i,g} \right)^T,$ with $\bvec{r}_{i,x} := \bvec{x}_i^m - \bvec{x}_i^{m+1} + \Delta t \left( \bvec{p}_i^m + m_i^{-1} \Delta t(\bvec{f}_i^m - \lambda_i\bvec{n}_i^m)/2  \right)$ and $r_{i,g} := g(\bvec{x}_i^{m+1}),$ is sufficiently small. In particular, we iterate until $\left\| \bvec{r}_i \right\| < \eta,$ with $\eta$ some small, positive tolerance. In our implementation, we use the infinity norm, which means that iterations continue until both $|r_{i,g}| < \eta$ and $\left\|\bvec{r}_{i,x}\right\| < \eta$ In section \ref{sec:results} we show that a tolerance of $10^{-6}$ produces sufficient energy conservation for a time step size of $0.0005\tau_{LJ},$ with $\tau_{LJ}$ the Lennard-Jones time unit defined below. The same iterative scheme is used for the update of the momentum, except we now denote the Lagrange multiplier with $\mu_i$ and the residual vector is now $\bvec{r}_i := \left( \bvec{r}_{i,p}, r_{i,np} \right)^T$ with $r_{i,np} := \bvec{n}_i^{m+1}\cdot \bvec{p}_i^{m+1},$ $\bvec{r}_{i,p} := \bvec{p}_{i}^{m+\half} - \bvec{p}_i^{m+1} + \Delta t \left( \bvec{f}_i^{m+1} - \mu_i \bvec{n}^{m+1} \right),$ and $\bvec{p}_i^{m+\half} = \bvec{p}_{i}^m + \Delta t \left( \bvec{f}_i^m - \lambda_i \bvec{n}_{i}^m \right).$ Note that at the time of the second iteration step, $\lambda_i$ and $\bvec{x}_{i+1},$ and thus $\bvec{p}_i^{m+\half}$ and $\bvec{n}_i^{m+1}$ are known.

\begin{algorithm}[t]
  \caption{RATTLE for particles on manifolds\label{alg:rattle}}
  \begin{algorithmic}
    \ForAll{$i$}
    \State $\lambda_i = 0,~\bvec{x}^{m+1}_{i}=\bvec{x}^m_i$
    \Repeat
    \State $\bvec{p}^{m+\half}_i = \bvec{p}^{m}_i + \frac{\Delta t}{2}\left( \bvec{f}_i^m - \lambda_i \bvec{n}_i^m \right)$
    \State $\bvec{r}_i = \begin{pmatrix}
      \bvec{x}_i^{m} - \bvec{x}_i^{m+1} + \frac{\Delta t}{m_i} \bvec{p}_i^{m+\half} \\
      g(\bvec{x}_i^{m+1})
    \end{pmatrix}$
    \State $\bvec{J}_i = \begin{pmatrix}
      -\bvec{I} & -\frac{(\Delta t)^2}{2 m_i} \bvec{n}_i^m \\
      \left( \bvec{n}_{i}^{m+1}\right)^T & 0
    \end{pmatrix}$
    \State $\begin{pmatrix} \bvec{x}_i^{m+1} \\ \lambda_i \end{pmatrix}
    = \begin{pmatrix} \bvec{x}_i^{m+1} \\ \lambda_i \end{pmatrix} - \bvec{J}^{-1}_i \bvec{r}_i$
    \Until{ $\left\|\bvec{r}_i\right\|<\eta$ }
    \EndFor
    \ForAll{$i$}
    \State $\mu_i=0,~\bvec{p}_i^{m+1} = \bvec{p}_i^{m+\half} + \frac{\Delta t}{2} \bvec{f}_i^{m+1}$
    \Repeat
    \State $\dot{\bvec{p}}_i^{m+1} = \bvec{f}_i^{m+1} - \mu_i \bvec{n}_i^{m+1}$
    \State  $\bvec{r}_i = \begin{pmatrix}
      \bvec{p}_i^{m+\half} - \bvec{p}_i^{m+1} + \frac{\Delta t}{2} \dot{\bvec{p}}_i^{m+1} \\
      m_i^{-1} \bvec{n}_{i}^{m+1}\cdot\bvec{p}_{i}^{m+1}
    \end{pmatrix}$
    \State $\bvec{J}_i = \begin{pmatrix}
      -\bvec{I} &  -\frac{\Delta t}{2} \bvec{n}_i^{m+1} \\
      m_i^{-1}\left( \bvec{n}_{i}^{m+1}\right)^T & 0
    \end{pmatrix}$
    \State $\begin{pmatrix} \bvec{p}_i^{m+1} \\ \mu_i \end{pmatrix}
    = \begin{pmatrix} \bvec{p}_i^{m+1} \\ \mu_i \end{pmatrix} - \bvec{J}^{-1}_i \bvec{r}_i$
    \Until{ $\left\|\bvec{r}_i\right\|<\eta$ }
    \EndFor
  \end{algorithmic}
\end{algorithm}

%Although we call Algorithm \ref{alg:rattle} RATTLE-M, it really is just the classical RATTLE scheme rewritten for the special case where the constraint for bonded particles $i$ and $j,$ $\sigma_{ij} = \left\| \bvec{x}_i - \bvec{x}_j \right\|^2 - d^2_{ij},$ with $d_{ij}$ the bond length, is replaced by the holonomic constraint of the curved surface for one particle, $g(\bvec{x}_i).$
{ Note that Algorithm \ref{alg:rattle} is just the classical RATTLE scheme written out for the special case there the constraint function only depends on the position of one particle.}
Writing it out illustrates more clearly two properties of RATTLE when it is applied to curved surfaces: The constraints for each particle only depend on the position, momentum and mass of that particle and not of the other particles, and the Jacobi matrices $\bvec{J}_i$ that appear in the RATTLE algorithm are computationally cheap to invert.
% In fact, the vector $\bvec{J}_i^{-1}\bvec{r}_i$ can be calculated analytically, which makes implementations of {RATTLE} very efficient.
{In fact, the equations involving the Jacobi matrix can be solved analytically, which makes implementations of RATTLE very efficient.}

\section{Results and discussion}
\label{sec:results}
In this section we first present some verifications of the presented method, assess its performance and finally discuss two examples where the method is applied. All results will be presented in so-called Lennard-Jones units, with length unit $\sigma,$ thermal energy unit $k_B T,$ mass unit $m_{LJ}$ and a resulting time unit of $\tau_{LJ} = \sigma  (m_{LJ}/k_B T)^{1/2}.$

\subsection{Verification}
\label{sec:verification}
As a first verification we measure how well the method conserves the total energy for a collection of 500 Lennard-Jones particles with no external forces acting on them. Energy drift tends to happen over long periods of time due to accumulation of round-off errors \cite{engle-2005}, but are acceptable as long as they are sufficiently small over the entire duration of the simulation.

For our test we implemented {Algorithm \ref{alg:rattle}} in LAMMPS \cite{plimpton-1995} and monitored the conservation of the total energy per particle for particles on four surfaces: a sphere, a torus, a plane and a cylinder.
% a sphere of radius $R=10\sigma,$ a plane of $L\times L$ with $L=30 \sigma$, a torus with large radius $R=10\sigma$ and small radius $r=4\sigma$ and a cylinder with radius $R=6 \sigma$ and length $L=50\sigma$.
For the plane we used periodic boundaries along x and y, and for the cylinder the boundary perpendicular to the cylinder's axis was also periodic. If we invoke a truncated and shifted Lennard-Jones potential with a time step size of $5\cdot10^{-4} \tau_{LJ},$ we observe a drift in the total energy per particle of at most 1.5\% over a time interval of $5\cdot10^6\tau_{LJ}$ ($10^{10}$ time steps) for particles confined to a sphere. If we instead use a linearly smoothed Lennard-Jones potential, we observe no noticeable energy drift over a time interval of $5 \cdot 10^{5} \tau_{LJ}$ ($10^9$ time steps), and the largest fluctuation is roughly $1.5\cdot10^{-3}\%,$ again for the spherical surface. The energy drifts for the cylinder, plane and torus were smaller for all cases.
These findings are reminiscent of those of Ref. \cite{toxvaerd-2012}, in which the energy conservation of a truncated shifted and a linearly smoothed Lennard-Jones potential were studied, revealing that linearly smoothed potentials are less prone to energy drift.

We explain that the energy drift for curved surfaces is larger by the presence of additional sources for round-off error, namely the iterative scheme used to solve the constrained equations of motion. We confirm this by again checking conservation of the total energy at a larger (less strict) tolerance of $\eta=10^{-4}$ for the same time step size of $5\cdot 10^{-4}\tau_{LJ},$ as well as for the original (stricter) tolerance of $\eta=10^{-6}$ but with a larger time step size $\Delta t = 0.005\tau_{LJ}.$ In both cases we observed a larger drift in the total energy for all surfaces mentioned before, with the sphere again having the largest energy drift. Based on the aforementioned, we conclude that {RATTLE} can sufficiently conserve the total energy of the system for practical applications.

For relatively short simulations the constraints on $\eta$ and $\Delta t$ are more lenient, as both the tests with $\eta = 10^{-4}$ and with $\Delta t = 0.005\tau_{LJ}$ did conserve energy well for $10^7$ time steps, which for most applications is more than sufficient. In this case the largest deviation was less $1.5\%$ for the particles on a sphere without any noticeable drift. For more details regarding energy conservation tests, we refer the reader to section SI 1.1. If stricter energy conservation is required, this can be achieved by a stricter tolerance and smaller time step size, at the expense of additional computing time.

After confirming that the method adequately conserves energy, we determined that by combining a simple Langevin thermostat with the constraint algorithm for three curved surfaces, Brownian motion is recovered. More specifically, we perform calculations with 2000 non-interacting particles constrained to a either sphere, a cylinder or a plane. For these surfaces, analytic expressions for the mean squared displacement $\left<\delta x^2\right>$ can be derived.
For the plane we have $\left<\delta x^2\right> = 4 Dt,$ for the cylinder $\left<\delta x^2\right> = 2 D t + 2R^2 ( 1 - \exp( -Dt/R^2) )$ and for the sphere $\left<\delta x^2\right> = 2R^2 \left( 1 - \exp(-2Dt/R^2) \right).$ The expressions are derived in SI 1.3. {RATTLE} in combination with the (already existing) LAMMPS implementation of the Langevin formalism described in Ref. \cite{gronbech-jensen-2013} reproduces these expressions. The largest root-mean-squared deviations from the analytical  expressions for $\left<\delta x^2\right>$ is $4.5 \sigma^2$ for the planar case, while the largest deviation is less than $10 \sigma^2,$ also for the planar case. After roughly $20 \tau_{LJ}$ the expressions converge to within $10\%$ of the theoretical values. More details{, including a pseudocode representation of the RATTLE update in combination with the Langevin thermostat,} are presented in SI 1.2.

{To verify that the Langevin thermostat properly keeps the temperature constant, we compute the temperature following the definition in \cite{leimkuhler-boek}, $k_B T = 2 K / N_{dof},$ where $K$ is the total kinetic energy of the particles and $N_{dof}$ the total number of degrees of freedom. $N_{dof}$ is the sum of the degrees of freedom of each particle, which is 3 for particles that move freely in 3D space but 2 for particles that are constrained to the curved surface. The temperature we measure according to this equation is indeed consistent with the temperature at which the thermostat is set.

The fact that RATTLE subtracts a component of the random force along the surface normal should not change the distribution generated in the 2D plane. If the three components of the random force are uncorrelated, subtracting the component along an arbitrary direction is the same as projecting onto an arbitrary plane. Because of the properties of normally distributed numbers, the random vector after a RATTLE correction is normally distributed in the plane tangent to the constraint function of the particle, and thus still has the correct distribution.}

We finally assessed the performance of the implementation of the algorithm in LAMMPS. We do so by considering a {larger number of Lennard-Jones particles, namely 10000 per processor core,} constrained to a sphere, and compare this with the same number of particles on a 2D plane with periodic boundary conditions. Constraining particles to a 2D plane is trivially achieved by not evaluating any position or velocity updates in the $z$-direction, so there is no overhead associated with this. Thus, comparing timings between a 2D plane and a sphere gives insight into the effective cost of the constraint algorithm. To make sure similar amounts of time are spent in force calculations, we ensure that both simulations are at an equal density. To assess the parallel scaling as well, we also varied the number of used processors, while keeping the number of particles per processor constant.

These benchmarks reveal that {RATTLE} is about a factor of 1.5 slower than the unconstrained update on a single node with 8 cores, up to a factor of 2 slower on 8 nodes with 8 cores. However, the parallel scaling of the algorithm is nearly as good as for the unconstrained system, as for 8 nodes, the parallel efficiency on of the unconstrained system was only about 1.2 times that of {RATTLE}. Therefore, although {RATTLE} requires more computing time since it needs to solve a constraint equation for each particle each time step, its parallel scaling is almost as strong as an unconstrained velocity Verlet scheme, making it excellent for simulating large numbers of particles. 
Finally we note that for very large systems, the performance of {RATTLE} could benefit from load balancing, in which an effort is made to assign each processor roughly the same number of {atoms to minimise} the idle time per processor. In our case load balancing did not make a difference, presumably due to the relatively uniform distribution of the particles over the sphere and the high cost of communication between the nodes of the computing cluster. More details relating to the benchmarks are presented in SI 2.

We now turn to two examples that illustrate the generality and flexibility of the method presented here. We first show in section \ref{sec:grana} how a complex curved surface can be built up from a combination of simple shapes in order to create a model of grana thylakoids connected by a lamella, and we then study how crowding affects diffusive processes on this shape. The second example in section \ref{sec:virus} illustrates how, by combining standard MD tools and the methods described here, one can study the self-assembly behaviour of complex particles on a spherical template. We use this model to determine how the bulkiness of a subdomain of the {particle} affects the shape of the self-assembled capsid, showing that very bulky {particles} generally lead to buckled capsids, rather than spherical ones.

%\subsection{The interplay between crowding and diffusion on curved membranes}
\subsection{Crowded diffusion on curved membranes}
\label{sec:grana}

%As noted in the Introduction, the lateral diffusion of proteins on curved membranes is one of the prime applications of our algorithm. In this example we showcase how our algorithm can capture the subtle interplay between crowding and membrane curvature. To do so we model two compartments, separated by a cylindrical structure, a generic motif present in grana thylakoid connected by a single lamella and dendritic spines at synapses. We measure the first passage time of proteins from one compartment to the other is affected by the surface density of crowders for two radii of the bridge (See figure x). This allows one to measure the average escape time which probes how the transport timescales of this process are influenced both due to the geometry and crowding.

As noted in the Introduction, the lateral diffusion of proteins on curved membranes is one of the prime applications of {RATTLE}. In this example we showcase how {RATTLE, applied to curved surfaces}, captures the interplay between crowding and membrane curvature. To do so we simulate diffusing particles on two compartments connected with a cylindrical ``bridge'', a generic motif present in grana thylakoid connected by a single lamella \cite{tremmel-2003} and dendritic spines at the synapse \cite{jackson-2011,straub-2012}. We measure how the first passage time of proteins from one compartment to the other is affected by the surface density of crowders for two bridge radii $R_b.$

\begin{figure}[htb!]
  \begin{center}
    % \begin{subfigure}[t]{0.50\textwidth}
    %   \includegraphics[width=\textwidth]{thylakoid_1}\caption{\label{fig:thyl-a}}
    % \end{subfigure}
    % \begin{subfigure}[t]{0.46\textwidth}
    %   \includegraphics[width=\textwidth]{thylakoid_2}\caption{\label{fig:thyl-b}}
    % \end{subfigure}
    % \begin{subfigure}[t]{0.48\textwidth}
    %   \includegraphics[width=\textwidth]{thylakoid_3}\caption{\label{fig:thyl-c}}
    % \end{subfigure}
    % \begin{subfigure}[t]{0.48\textwidth}
    %   \includegraphics[width=\textwidth]{thylakoid_4}\caption{\label{fig:thyl-d}}
    % \end{subfigure}
    \includegraphics[width=0.9\textwidth]{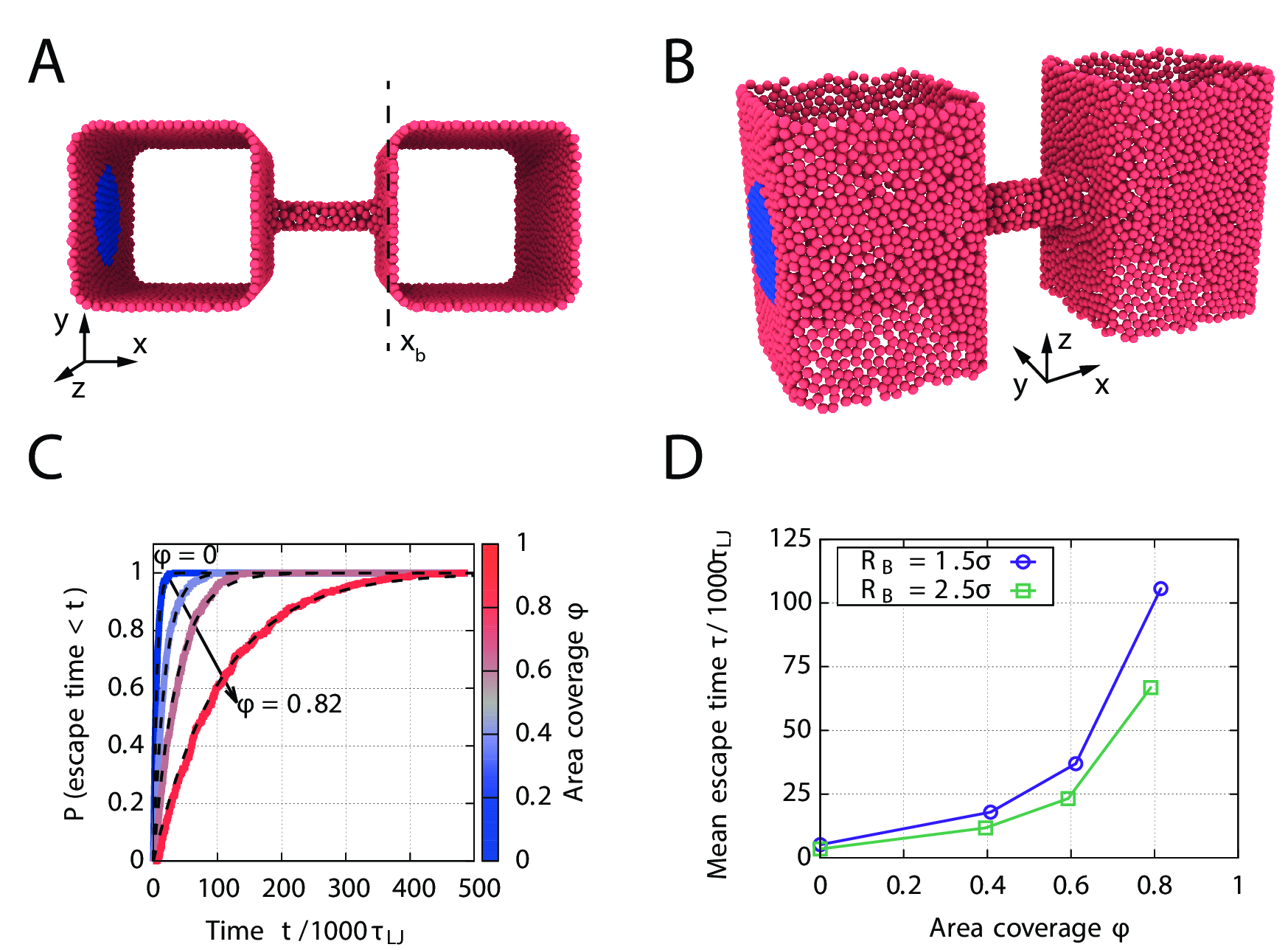}
    \caption{(A,B): Initial configuration of particles with diameter $2^{1/6}\sigma$ on two blocks connected by a cylindrical bridge at surface area coverage $\phi=0.82$ for $R_b = 1.5\sigma.$ Black lines indicate periodic 
      boundaries. We place about 100 tracer particles (blue) on one block. (B). (C): Empirical cumulative probability density functions of the time it takes for the tracers to reach $x \geq x_b$ for the first time as function of $\phi$ (solid) and exponential distributions with the same mean (dashed). (D): Average escape times for all $\phi$ considered for two bridge radii $R_b = 1.5\sigma$ (squares) and $R_b = 2.5\sigma$ (circles). \label{fig:thylakoid-escape-times}}
  \end{center}
\end{figure}

To measure how crowding influences escape times, we place tracer particles on the back of one compartment and fill the rest of the surface with a varying surface area density of crowders. All particles interact with each other through a Lennard-Jones potential truncated and shifted at $r = 2^{1/6}\sigma.$ We associate this length scale with the effective diameter of the particles $d_0.$ By truncating and shifting at this distance the particles repel each other when close but do not attract over longer distances. We apply a Langevin thermostat to both the crowders and the tracer particle to make them undergo Brownian motion. We then determine the escape time $\tau_e$ it takes for the tracer particle to reach the other compartment, mathematically expressed as $x\geq x_{b},$ with $x_{b}$ the x-coordinate at which the bridge connects to the other compartment (see figure \ref{fig:thylakoid-escape-times}A).

All times in this section are now expressed as multiples of the damping time in the Langevin equation, which was {put equal} to unity. This damping time is the time it takes for velocity autocorrelation effects to decay to a fraction of $1/e$ of the zero-time value. We measure the escape times for 5 different random seeds for each area coverage. This area coverage is determined by assigning to the particles an effective area they cover, $A_p = \pi d_0^2/4,$ with $d_0$ the distance at which the interaction potential is truncated. The total area the particles cover is then given by $N A_p,$ and the area fraction $\phi$ can be obtained by dividing this total area with the area of the total thylakoid surface $A_t,$ $\phi := N A_p / A_t.$ Obtaining $A_t$ is not difficult but tedious because of the many parts of which the surface is constructed, so we leave the derivation for SI 3.1. To get a feeling for how crowding affects the diffusive behaviour, we also sample escape times for non-interacting particles, which we associate with the dilute limit $\phi = 0.$

From these data we can determine the average escape time, which should approximate the mean escape time, as well as the underlying distribution. This way we determine how an increasing density, and hence an increased effect of crowding, affects the mean escape time. In figures \ref{fig:thylakoid-escape-times}A and \ref{fig:thylakoid-escape-times}B we show the initial setup of the system considered. Note that in this case, the constraint function is defined {piece-wise}. {RATTLE} allows for this as long as the surface normals are {continuous} along the edges of the subdomains. This provides a lot of flexibility, as intricate surfaces can be decomposed into subdomains with simple constraint functions.
{Note that if the definition for the surface normal is not differentiable across the subdomains, this can degrade the performance of the algorithm. Hence, it is best to define the holonomic constraints in such a way that the surface normal is differentiable across the domains.}
The complete definition of the constraint function for this surface is given in SI 3.1.

In figure \ref{fig:thylakoid-escape-times}C we show, for a few packing fractions, the typically observed distribution of escape times, as well as an exponential distribution with the same mean. This figure suggests that the escape times for a fixed density are approximately exponentially distributed. Furthermore, in figure \ref{fig:thylakoid-escape-times}D the influence of the crowder density on the escape times is clearly seen. It is however important to distinguish between effects caused by a smaller bridge and connector on the one hand, and a higher density on the other hand. Therefore, we turn first to the results for the non-interacting case $\phi=0.$

For $\phi=0,$ the escape times for the smaller bridge radius, $R_b = 1.5\sigma,$ is about 1.125 larger than for $R_b=2.5\sigma.$ This increase is caused by the fact that, in order to reach the other block, particles first need to find the connector \cite{holcman-2014,kusters-2013}. Thus, a smaller bridge already leads to increased escape times. For higher surface area fractions, however, the difference between the escape times becomes larger.
This is because for a smaller cylinder, the point where particles can move only through collective motions is reached sooner, as was shown for a cylinder of varying radius in Ref. \cite{kusters-2015}. Thus, the escape times for particles on curved surfaces are influenced drastically by two things: The probability of finding the ``exit'' on the one hand, and the effects of crowding on the other hand.

The important contribution of a scheme like {RATTLE} is that now particle-particle and particle-crowder interactions can be taken into account explicitly. This is in contrast to cited works, as in {Ref.} \cite{holcman-20142} only crowding effects due to immobilized, inaccessible spherical regions are considered. In {Ref.} \cite{kusters-2013,kusters-2014} no crowding effects are considered at all. Finally, in {Ref.} \cite{tremmel-2003} a Monte Carlo simulation of tracers and immobilized crowders on a lattice is performed, with only crowding effects between tracers and obstacles included. In principle, they could include tracer-tracer interactions by treating a tracer-occupied lattice cite as inaccessible as well, but given that the tracers can only move on a lattice, the question is how realistic this model would be. In contrast, with {RATTLE} the particle-particle interactions are included in a lattice-free manner, and thus allows for realistic collective motions.

\subsection{Virus capsid self-assembly}
\label{sec:virus}
As another example we present here how {RATTLE} can be used in combination with bond and angle potentials to model complex molecules of which parts are constrained to a surface. In particular, we consider conical particles, which model capsomeres in viruses, inspired by the simulations of Chen \emph{et al.} \cite{chen-2007-1,chen-2007-2} and Yu \emph{et al.} \cite{hagan-2015}. For conical particles in free space, Chen found that for certain ranges of parameters, these particles robustly self-assemble into {icosahedral} structures. Yu and Hagan \cite{hagan-2015} suggest the use of cylindrical particles to more accurately capture the effective shape of virus capsid proteins, \eg{}, for HIV, of which the capsid proteins are in shape closer to rods than to spheres.

\begin{figure}[htb!]
  \begin{center}
    % \subcaptionbox{\label{fig:capsomere-model-lean}}{
    %   \includegraphics[height=4cm]{virus_1}
    % }
    % \subcaptionbox{\label{fig:virus-shrink1}}{
    %   \includegraphics[height=4cm]{virus_2}
    % }
    % \subcaptionbox{\label{fig:virus-shrink2}}{
    %   \includegraphics[height=4cm]{virus_3}
    % }
    % \subcaptionbox{\label{fig:virus-buckle}}{
    %   \includegraphics[height=4cm]{virus_4}
    % }
    % \subcaptionbox{\label{fig:virus-radii}}{
    %   \includegraphics[height=4.4cm]{virus_5}
    % }
    \includegraphics[width=0.9\textwidth]{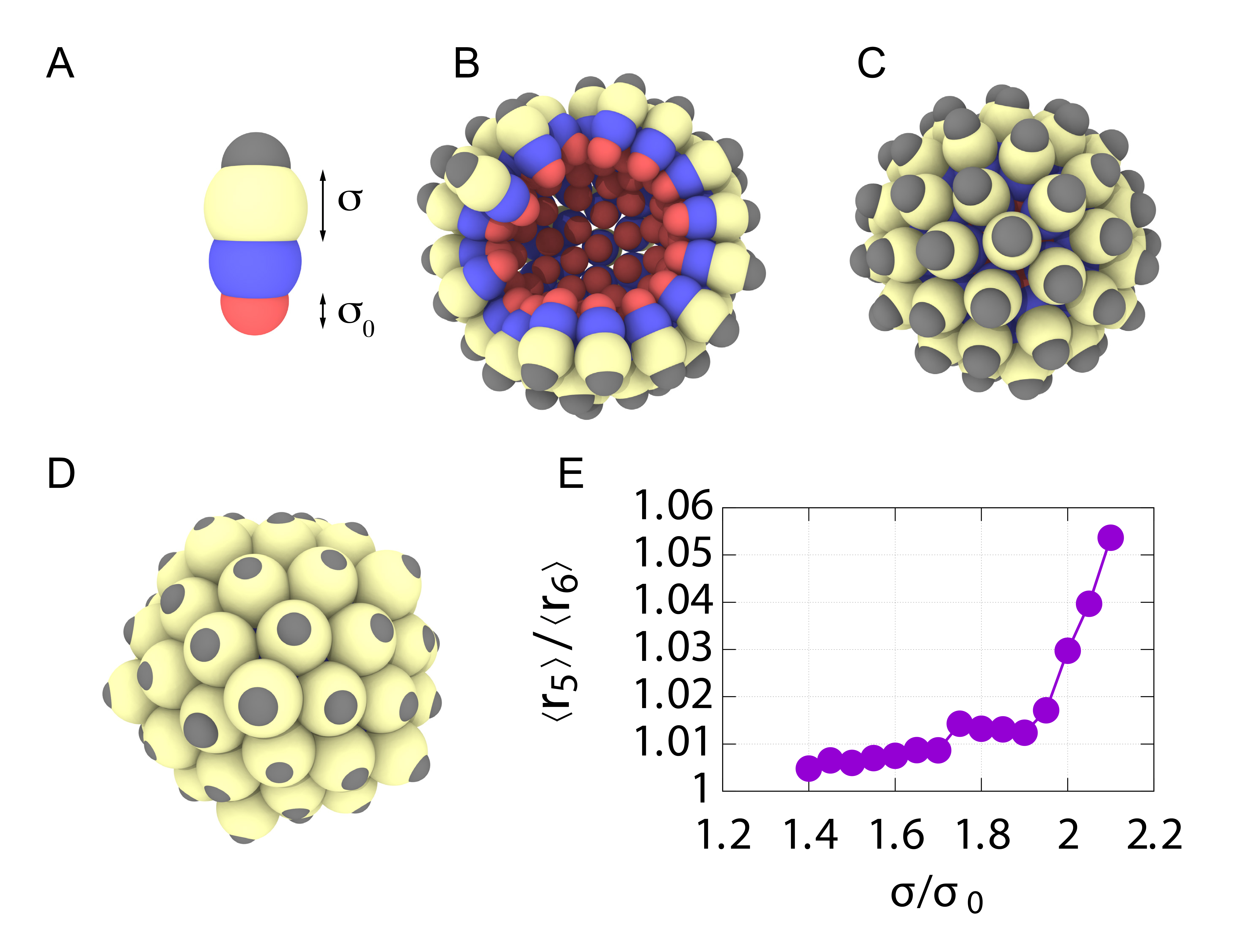}
    \caption{(A): The capsomere model. The effective size of the beige bead, $\sigma,$ is varied from $1.4\sigma_0$ to $2.1\sigma_0,$ with $\sigma_0$ the size of the red bead. The beads are held together by harmonic springs. (B): Capsomeres with $\sigma = 1.4\sigma_0$ on a spherical surface. The radius of the surface is not the equilibrium radius and the capsomeres form an incomplete capsid. (C): Equilibrium structure for $\sigma = 1.4\sigma_0.$ (D): Buckled equilibrium capsid for $\sigma=2.1\sigma_0.$ (E): Equilibrium averages of the distance from origin to beige beads that have five neighbours ($\left<r_5\right>$) or six ($\left<r_6\right>$). Upon buckling, the beige beads are pushed outward, manifesting itself in a larger ratio $\left<r_5\right>/\left<r_6\right>.$   \label{fig:virus-example}}
  \end{center}
\end{figure}

Our capsomeres consist of four beads, as illustrated in figures \ref{fig:virus-example}A. We constrain the red bead to a spherical surface, representing the RNA \cite{cadena-nava-2012} or a nanoparticle \cite{chen-2006} to which the capsomere binds. The beads in each capsomere are connected with harmonic springs, but are otherwise free to explore all of space. We also employ an {angular} potential to prevent the cone from bending.
Beige and blue beads in different capsomeres attract through a Lennard-Jones potential, while grey and red beads are purely repulsive. By changing the effective size of the beige bead, $\sigma,$ one can determine how the geometric properties of capsomeres influence the self-assembly behaviour. More specifically, we consider values for $\sigma$ between $1.4\sigma_0$ and $2.1 \sigma_0,$ where $\sigma_0$ is the size of the red bead. The interaction strength of the Lennard-Jones potential {is} $4 k_B T$ for all beads.

For all values of $\sigma$ considered, the capsomeres assemble into an icosahedral capsid, meaning that there are capsomeres with five nearest neighbours and capsomeres with six nearest neighbours, which we call pentamers and hexamers, respectively. We find that above a critical value of the size $\sigma,$ the outer parts of the capsomere become too large for the template and the capsid takes on a buckled configuration rather than a spherical one, in which the pentamers are pushed outwards further than the hexamers. The excess strain due to the nonconforming capsomere size are thus concentrated on the pentamers. This is reminiscent of Ref. \cite{lidmar-2003}, in which it was found that virus capsids, when sufficiently large, release their elastic strain also by buckling the twelve five-fold disclinations, giving rise to aspherical, faceted particles very similar to the ones we observe. 

For this example we are only interested in equilibrium properties and not dynamics, so we invoke a Nosé-Hoover thermostat as discussed in \cite{martyna-1994} instead of a Langevin thermostat.
{Note that for purely harmonic systems, the Nosé-Hoover thermostat can have ergodicity problems, as explained in, \emph{e.g}, Refs. \cite{hoover-1985,legoll-2006}. In this example, however, the Lennard-Jones interactions are anharmonic, and thus the Nosé-Hoover thermostat should properly sample the canonical ensemble.}
For more details about the simulation setup, we refer the reader to SI 3.2. In order to identify the preferred capsid  size, we slowly shrink the spherical template and determine for which radius $R$ the potential energy is minimal, that we denote $R^*.$ We then sample the equilibrium positions of the beads for this radius $R^*.$

In figures \ref{fig:virus-example}B and \ref{fig:virus-example}C we show two snapshots of a self-assembling capsid on a shrinking template. In figure \ref{fig:virus-example}B, the template is larger than the preferred radius of the capsid, which, in combination with the attractive interactions between the blue and beige beads, leads to a hole on one side of the template. At a later time, the template has sufficiently shrunk, and the hole closes.

At the values of $\sigma$ we considered, we observe either nearly spherical capsids, like the one illustrated in \ref{fig:virus-example}C, or buckled capsids like the one illustrated in figure \ref{fig:virus-example}D. The buckling manifests itself in the protrusion of the pentamers from the otherwise spherical capsid, so we quantify buckling by the ratio of the distance from the origin to the beige beads in the aforementioned pentamers and hexamers, $r_5$ and $r_6$ respectively. Hence, the ratio $r_5/r_6$ is a measure for how buckled the capsid is. In figure \ref{fig:virus-example}E we show equilibrium averages of this quantity as a function of $\sigma$ at the final radius of the template $R^*.$ These indicate that significant buckling only takes place for $\sigma > 1.9\sigma_0,$ indicating that only for a very large mismatch between capsomere and template buckling takes place.

\section{Conclusions}
\label{sec:conclusion}
Constraining particles with the RATTLE algorithm provides a useful and flexible tool to study motion of particles on manifolds for, \eg{}, their diffusive properties and the equilibrium structures they assume.
The {RATTLE} variant proposed here was implemented as a module for LAMMPS, allowing, without any modifications, the use of many features, including but not limited to:
a wealth of interaction potentials, parallel tempering/replica exchange molecular dynamics, and bond/angle potentials to model bead-and-spring polymer and protein models.

We illustrated how the scheme can give insight into diffusion dynamics of simple particles on complex surfaces. Furthermore we showed how {RATTLE} can be used to model self-assembly of complex particles on a spherical template, a model especially relevant for viruses. Another possibility not illustrated in this work would be to apply the constraints to only some of the particles or proteins in the simulation box in order to model the interplay between particles diffusing along a curved surface and the crowded solvent surrounding it.

In terms of performance, {RATTLE} is only about a factor of 1.5 to 2 slower than a two-dimensional unconstrained velocity Verlet scheme at a similar density, with some variation depending on the communication speed between nodes of the used computing cluster and the computational cost of the constraint functions. The parallel performance might be improved with active load balancing, which attempts to keep the number of particles per processor constant in an attempt to minimise the idle time for each processor. In our case this did not matter, however, probably because the performance bottleneck of our computing cluster was clearly the parallel communication.

The performance and flexibility mentioned above make {RATTLE} an appropriate scheme to study dynamics of complex particles on curved surfaces, especially for larger systems due to the trivial parallelisation of the scheme.

\section*{Supporting Citations}
References \cite{lelievre,vandeneijnden-2006,lelievre-2012,frenkel-boek} appear in the Supporting Material.

\section{Acknowledgements}
We thank Paul van der Schoot for proof-reading and Wouter Ellenbroek for suggesting a good performance benchmark. S.P. acknowledges the HFSP for funding under grant RGP0017/2012 and R. K. acknowledges FOM for funds from the Netherlands Organization for Scientific Research (NWO-FOM) within the program ‘‘Barriers in the Brain: the Molecular Physics of Learning and Memory’’ (No. FOME1012M).

%\singlespacing

\def\bibfont{\footnotesize}

\clearpage{}

\renewcommand{\cite}[1]{\citep{#1}}

\setcounter{section}{0}
\setcounter{figure}{0}
\setcounter{algorithm}{0}
\setcounter{table}{0}
\setcounter{equation}{0}

\renewcommand{\theequation}{S\arabic{equation}}
\renewcommand{\thealgorithm}{S\arabic{algorithm}}
\renewcommand{\thefigure}{S\arabic{figure}}
\renewcommand{\thesection}{Section S\arabic{section}}
\renewcommand{\thesubfigure}{\Alph{subfigure}}
\renewcommand{\thetable}{S\arabic{table}}

\section*{Supporting information: A method for molecular dynamics on curved surfaces}

\section{Verification results}
This section contains details about the two verifications mentioned in section 3.1. We first show that {RATTLE} conserved energy as good as an unconstrained Verlet integration scheme by comparing it to known results from ref \cite{toxvaerd-2012}. We then illustrate that a simple Langevin thermostat applied to particles constrained on curved surfaces reproduces the expected Brownian motion. To this end, we implemented {RATTLE} as a module for LAMMPS \cite{plimpton-1995}, which already contained a module for the aforementioned Langevin dynamics, which adds every time step a damping term and random force to the forces acting on each particle.

\subsection{Energy conservation}
We firstly checked how well {RATTLE} conserves the total energy of the system $\mathcal{H} := K + U,$ with $K$ and $U$ the total kinetic and potential energy, respectively. To do so, we constrain $N=500$ particles to a 2D plane, cylinder, torus and sphere, and integrate the system over a long time. We chose as interaction potentials both a truncated shifted Lennard-Jones potential $V_{LJ}$ and a linearly smoothed truncated Lennard-Jones potential $V_{LJ}^*.$ Their respective formulas are:
\begin{align}
  \phi(r) =& 4\epsilon\left[\left(\frac{\sigma}{r}\right)^{12} - \left(\frac{\sigma}{r}\right)^6 \right], \qquad V_{LJ} := \begin{cases}
    \phi(r) - \phi(r_c) &\quad\text{if }r < r_c,\\
    0&\quad\text{otherwise.} \label{eqn:LJ-trunc-shift}
  \end{cases} \\
  V_{LJ}^* :=& \begin{cases}
    \phi(r) - \phi(r_c) + (r-r_c)\frac{d \phi}{d r}(r_c) &\quad \text{if } r < r_c\\
    0&\quad\text{otherwise.} \notag % \label{eqn:LJ-smooth-linear}
  \end{cases}
\end{align}
In the case of $V_{LJ},$ the energy is continuous at the cut-off distance $r_c$ but the forces are not, while in the case of $V_{LJ}^*$ both the energy and the forces are continous at $r_c.$ Because of this, linearly smoothed potentials tend to conserve energy better in unconstrained simulations \cite{toxvaerd-2012}.

\begin{figure}
  \begin{center}
    \includegraphics[width=0.8\textwidth]{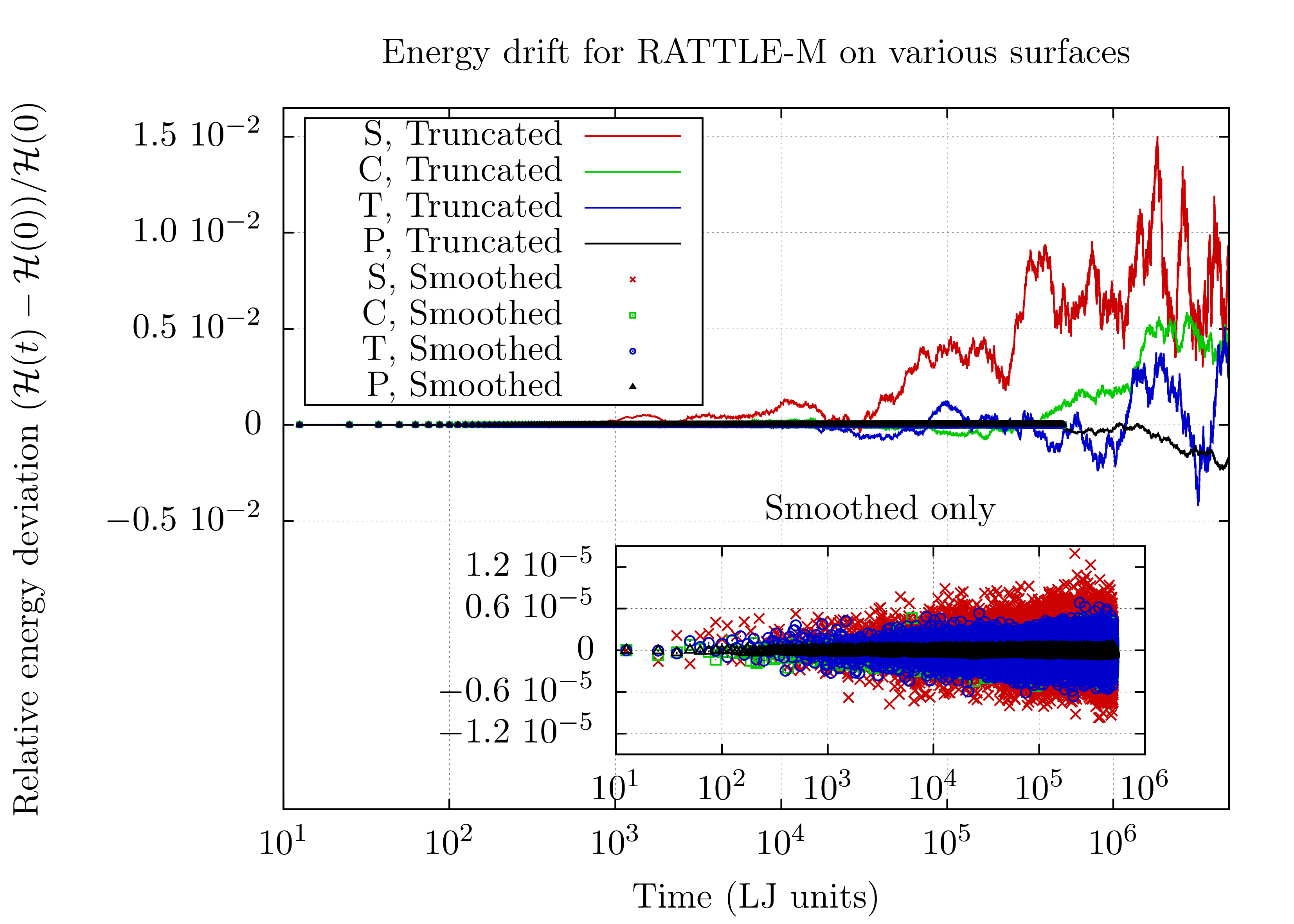}
    \caption{Energy conservation for a system of 500 particles constrained to a sphere (S, red), cylinder (C, green), torus (T, blue) and a 2D plane (P, black) using {RATTLE}, with both a truncated shifted Lennard-Jones (Truncated) and a truncated linearly smoothed Lennard-Jones (Smoothed) potential. We used a time step size $\Delta t = 0.0005\tau_{LJ}.$ For the truncated potential there is a noticeable drift in total energy after a time interval of $1000\tau_{LJ}$ (2 million time steps), although it is less than 2\%. For the smoothed potential there are fluctuations about the total energy in the order of $10^{-3} \%$, but there is no noticeable drift from the total energy. \label{fig:energy-conservation}}
  \end{center}
\end{figure}

In figure \ref{fig:energy-conservation} we show the evolution of the system's Hamiltonian relative to its initial value, over a simulation length of $10^6$ Lennard-Jones time units with a time step of $\Delta t = 0.0005.$ This clearly shows that for the truncated Lennard-Jones system, the total energy of the system deviates noticeably but not significantly for very long simulations. For the smoothed potential we see that, over this simulation length, the total energy fluctuates but does not noticeably drift, implying that {RATTLE} conserves energy sufficiently well, especially when the forces are continuous at the potential cut-off. These findings remind one of observations for unconstrained systems presented by Toxvaerd \cite{toxvaerd-2012} in which a smoothed truncated potential also results in much better energy conservation.

\subsection{Diffusion}
After verifying proper energy conservation, we determined that by combining the Gr{{\o}}nbech-Jensen/Farago formulation for Langevin dynamics \cite{gronbech-jensen-2013} with {RATTLE}, Brownian dynamics on manifolds can be generated. {See Algorithm \ref{alg:rattle-plus-langevin} for a pseudocode representation of the complete time integration step obtained this way.}

{
\begin{algorithm}[t]
  \caption{RATTLE with Gr\/{o}nbech-Jensen/Farago Langevin formalism\label{alg:rattle-plus-langevin}}
  \begin{algorithmic}
    \ForAll{$i$}
    \State Perform the same position update as in Algorithm 1.
    \EndFor
    \ForAll{$i$}
    \State Compute new forces $\bvec{f}_{i}^{m+1}$ from $-\nabla V$
    \State Draw $r_x, r_y$ and $r_z$ from a normal distribution with $\mu=0, \sigma^2=1$
    \State Generate new random force vector $\bvec{f}^{n+1}_r = \sqrt{2m_i k_B T / \tau  \Delta t} (r_x,r_y,r_z)^T$
    \State Combine forces: $\bvec{f}_{i}^{m+1} = \frac{1}{1 + \Delta t/2\tau }\left[ \bvec{f}_{i}^{m+1} - \frac{1}{\tau}\bvec{p}^{m+\half}+ \half\left(\bvec{f}_r^{n+1} + \bvec{f}_r^n \right)\right]$
    \EndFor
    \ForAll{$i$}
    \State Perform the same momentum update as in Algorithm 1.
    \EndFor
  \end{algorithmic}
\end{algorithm}
}

\begin{figure}[htb]
  \begin{center}
    \includegraphics[width=0.8\textwidth]{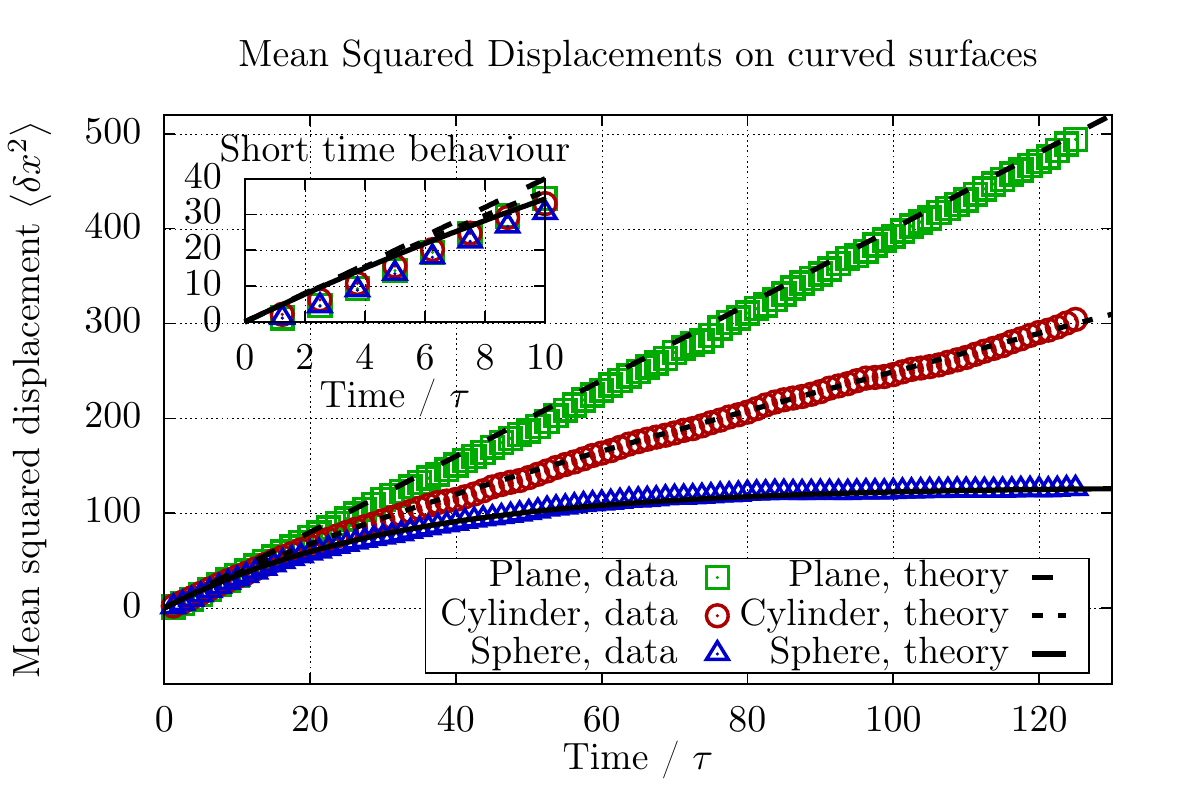}
    \caption{Mean squared displacement of 2000 non-interacting particles subject to a Langevin thermostat and constrained to a 2D plane (green squares), a cylinder (red circles) and a sphere (blue triangles). The dashed lines represent theoretical results (see section \ref{sec:msd-theory} for derivations). The damping time in the Langevin equation was set to correspond to a diffusivity of $1 \sigma_{LJ}^2/\tau_{LJ}.$ For short time scales, the particles do not ``feel'' that they are on a curved surface yet, and in all cases the MSD grows linearly in time (inset).\label{fig:diffusion-surfaces}}
  \end{center}
\end{figure}

This implementation parallelises trivially. Its performance is benchmarked in S.3.3, where we find that the performance scales linearly with the number of cores for sufficiently large systems on a single node and that an update with the {RATTLE} implementation in LAMMPS is about a factor of 1.5 slower than an unconstrained velocity Verlet update in 2D at an equal density. The parallel scaling is as effective as the standard velocity Verlet implementation of LAMMPS. The details about the setup of the simulations are given in S.3.2, the results are illustrated in figure \ref{fig:diffusion-surfaces} where it is shown that the combination of {RATTLE} and a Langevin thermostat correctly reproduces theoretical expressions for the mean squared displacement (MSD). These simulations were done for $N=2000$ non-interacting particles. Thus, the combination of a Langevin thermostat using the Gr{{\o}}nbech-Jensen/Farago formulation in combination with {RATTLE} reproduces the expected diffusive behaviour and can be used to study diffusion on more complex curved surfaces.

We do want to point out that, while the Langevin scheme used here properly generates Brownian motion, there are more advanced methods available (see, e.g., ref. \cite{lelievre,vandeneijnden-2006,lelievre-2012}) that can also be combined with {RATTLE}. However, demonstrating those schemes goes beyond the main goal of this paper, which is to show that by combining standard tools from molecular dynamics, it is possible to efficiently simulate large scale coarse grained models of particles confined to curved surfaces, in particular diffusion along curved membranes in biological systems.

Finally, note that if one is not actually interested in dynamics, the Langevin thermostat can be replaced with a Nosé-Hoover type thermostat \cite{frenkel-boek,martyna-1994} to determine the equilibrium properties of the system at constant temperature.

\subsection{Derivation of mean squared displacement}
\label{sec:msd-theory}
This section describes how the expressions for MSDs on the curved surfaces considered in figure \ref{fig:diffusion-surfaces} can be derived.

\subsubsection{Cylinder}
In the case of a cylinder of radius $R$ along the $z$-axis, the diffusion in the $z$-axis remains unaltered. Thus the MSD can be written as $\left< (\delta x)^2 \right> = 2Dt + \left< (\delta x_c)^2 \right>,$ with $\delta x_c$ representing the displacements in the x and y direction. This term is simply the MSD for particles on a circle, which can be derived by switching to polar coordinates. To find the probability of finding a particle at $(R\cos\phi, R\sin\phi)$ given that at $t = 0$ it was at $(x,y) = (R,0)$ can be found by solving the diffusion equation in polar coordinates:
\begin{equation*}
  \frac{\partial p}{\partial t} = \frac{D}{R^2}\frac{\partial^2 p }{\partial \phi^2}
\end{equation*}
% Rearranging and separating variables using $p(\phi,t) = \Phi(\phi)T(t)$ leads to
%\begin{equation*}
%  \frac{D}{R^2}\frac{\Phi''}{\Phi} = \frac{T'}{T} := -\lambda^2,
%\end{equation*}
%where $\lambda$ is an as of yet unknown parameter. From this equation we see that $T = A\exp(-\lambda^2t),$ with $A$ another arbitrary constant, that can be set to $1$ without loss of generality. Solving for $\Phi$ leads to
%\begin{equation*}
%  \Phi(\phi) = A \cos\left( \lambda R \phi / \sqrt{D} \right) + B \sin \left( \lambda R \phi / \sqrt{D}\right)
%\end{equation*}
%Since $p$ should be periodic in $\phi$ over $2\pi,$ it must hold that $\Phi(0) = \Phi(2\pi),$ which leads to
%\begin{equation*}
%  \cos \left( 2\pi \lambda R / \sqrt{D} \right) = 1,\qquad \sin\left( 2\pi \lambda R /\sqrt{D} \right) = 0
%\end{equation*}
%This can be achieved if $\lambda = n \sqrt{D} / R,$ with $n$ an integer. This all combined leads to
A general solution to this equation is given by
\begin{equation*}
  p(\phi,t) = \sum_{n=0}^\infty B_n \cos( n\phi ) \exp \left[ -\frac{n^2 D t}{R^2} \right] = B_0 + \sum_{n=1}^\infty B_n \cos( n\phi ) \exp \left[ -\frac{n^2 D t}{R^2} \right]
\end{equation*}
Normalizing this probability distribution leads to $B_0 = 1/(2\pi)$ and no information about $B_{n > 0}.$ To find those coefficients, we apply ``Fourier's trick'' to the initial condition $p(\phi,t=0) = \delta(\phi),$ which corresponds to a particle located at $x = R, y = 0$ at $t=0.$ This leads to
\begin{align*}
  \int_{\phi=0}^{2\pi} &\left[ 1 + \sum_{n=1}^\infty B_n \cos( n \phi ) \right] \cos(m \phi)d\phi \\
  =& 0 + \sum_{n=1}^\infty B_n \int_{\phi=0}^{2\pi} \cos(n\phi)\cos(m\phi)d\phi = \sum_{n=1}^\infty B_n \int_{\phi=0}^{2\pi} \cos^2(m\phi) \delta_{mn} d\phi \\
  =& \sum_{n=1}^\infty B_n \delta_{mn} \int_{\phi=0}^{2\pi}\left( \half + \half \cos(2m\phi) \right) d\phi = \sum_{n=1}^\infty \delta_{mn} B_n \pi = B_m \pi
\end{align*}
This expression should be equal to $\cos( m\phi )$ integrated over $p(\phi,t=0) = \delta(\phi),$ which is just $\cos(0) = 1,$ so we find that $B_{n>0} = 1 / \pi,$ so we find for the probability $p(\phi,t)$ of finding a particle at angle $\phi$ at a given time $t,$ given that it was at $\phi=0$ at time $t=0,$ that
\begin{equation}
  p(\phi,t) = \frac{1}{2\pi}\left[ 1 + 2 \sum_{n=1}^\infty \cos(n\phi)\exp\left( -\frac{n^2Dt}{R^2} \right) \right]
  \label{eqn:prob-cyl}
\end{equation}
To find the MSD, we now just calculate the squared distance $\delta x_c^2$ from $(R,0)$ to an arbitrary point $(R\cos\phi,R\sin\phi),$ and use $p(\phi,t)$ to find the expectation value of that quantity.
It is trivial to show that $\delta x_c^2 = 2R^2(1 - \cos(\phi)),$ and thus we find for the MSD
\begin{align*}
  \left< \delta x_c^2 \right> =& \int_{\phi=0}^{2\pi} \left[2R^2(1 - \cos\phi)\right]\frac{1}{2\pi}\left[1 + 2\sum_{n=1}^\infty \cos(n\phi)\exp\left(-\frac{n^2Dt}{R^2}\right)\right]d\phi \\
  =& \frac{R^2}{\pi}\int_{\phi=0}^{2\pi} (1 - \cos\phi)\left[1 + 2\sum_{n=1}^\infty \cos(n\phi)\exp\left(-\frac{n^2Dt}{R^2}\right)\right]d\phi \\
  =& \frac{R^2}{\pi}\int_{\phi=0}^{2\pi} d\phi - \frac{2R^2}{\pi}\sum_{n=1}^\infty  \int_{\phi=0}^{2\pi}\cos\phi\cos(n\phi) \exp\left(-\frac{n^2Dt}{R^2}\right)d\phi \\
  =& 2R^2\left[1 - \frac{1}{\pi}\sum_{n=1}^\infty \delta_{n1} e^{-n^2Dt/R^2} \int_{\phi=0}^{2\pi}\cos^2(\phi)d\phi\right]  = 2R^2\left[1 - e^{-Dt/R^2} \right].
\end{align*}
The total MSD thus is given by
\begin{equation}
  2Dt + 2R^2( 1 - e^{-Dt/R^2} )
  \label{eqn:msd-cyl}
\end{equation}
Note that in the limit of $t\rightarrow 0,~\left<\delta x_c^2\right> = 2R^2(1 - (1 - Dt/R^2)) = 2Dt,$ so for short times, where the particles do not ``feel'' the geometric confinement, the MSD is just $4Dt,$ like in a 2D plane.

\subsubsection{Sphere}
For a sphere a similar strategy as for the cylinder can be followed. We now start with a particle located in the ``north pole:'' $(0,0,R).$ The MSD can then be obtained from the probability of finding a particle at a location $(R\sin\theta\cos\phi,R\sin\theta\sin\phi,R\cos\theta)$ at some time $t,$ which again follows from the diffusion equation. Assuming radial symmetry, we only need to know what the probability is to find a particle at a polar angle $\theta$ at some time $t,$ so the diffusion equation becomes
\begin{equation*}
  \frac{\partial p}{\partial t} = \frac{D}{R^2 \sin\theta}\frac{\partial}{\partial \theta}\left(\sin\theta\frac{\partial p}{\partial \theta}\right).
\end{equation*}
This equation has general solutions in the form of spherical harmonics:
\begin{equation*}
  p(\theta,t) = \sum_{l=0}^\infty B_l Y_{l0}(\theta) \exp\left[-\frac{l(l+1)Dt}{R^2}\right],\qquad Y_{l0}(\theta) := \sqrt{\frac{2l+1}{4\pi}}P_l(\cos \theta),
\end{equation*}
where $P_l(x)$ is the $l$th Legendre polynomial. Applying ``Fourier's trick'' again to the initial condition leads to
\begin{align*}
  &\int_{\theta=0}^\pi Y_{k0}(\theta) \sum_{l=0}^\infty B_l Y_{l0}(\theta) \sin \theta d\theta = \sum_{l=0}^\infty B_l \int_{\theta=0}^\pi Y_{k0} Y_{l0} \sin \theta d\theta \\
  = &\sum_{l=0}^\infty B_l \delta_{lk} \int_{\theta=0}^\pi Y_{k0}^2 \sin\theta d\theta = \sum_{l=0}^\infty B_l \delta_{lk}  \int_{\theta=0}^\pi \frac{2l+1}{4\pi} \left[ P_k(\cos\theta) \right]^2 \sin \theta d\theta \\
  = &\sum_{l=0}^\infty B_l \delta_{lk} \frac{2l+1}{4\pi} \int_{x=-1}^1  \left[ P_k(x)\right]^2dx = \sum_{l=0}^\infty B_l \delta_{lk} \frac{2l+1}{4\pi} \frac{2}{2k+1} = \frac{B_k}{2\pi}
\end{align*}
Integrating the form of the initial condition, $p(\theta,t=0) = \delta(\theta)/\sin\theta$ leads to
\begin{align*}
  \int_{\theta=0}^\pi Y_{k0}(\theta) (\delta(\theta) / \sin\theta) \sin \theta d\theta = Y_{k0}(0) = \sqrt{ \frac{2k+1}{4\pi}} P_k(\cos 0) = \sqrt{\frac{2k+1}{4\pi}}
\end{align*}
Thus, $B_k = \sqrt{\pi(2k+1)}$ and the solution to the diffusion equation becomes
\begin{align*}
  p(\theta,t) &= \sum_{l=0}^\infty \sqrt{ (2l+1)\pi} Y_{l0}(\theta) \exp\left[-\frac{l(l+1)Dt}{R^2}\right] \\
  &= \sum_{l=0}^\infty \frac{2l+1}{2}P_l(\cos\theta) \exp\left[-\frac{l(l+1)Dt}{R^2}\right] 
\end{align*}
The distance from $(0,0,R)$ to a point on the sphere is now given by $2R^2(1 - \cos\theta).$ To exploit orthonormality of the Legendre polynomials in the integral later, we associate with $1$ and $\cos \theta$ the zeroth and first Legendre polynomials: $1 = P_0(\cos\theta)$ and $ \cos \theta =  P_1(\cos\theta).$
The MSD thus becomes
\begin{align*}
  \left<\delta x^2\right> =& \int_{\theta=0}^\pi 2R^2( 1 - \cos\theta ) \sum_{l=0}^\infty \frac{2l+1}{2}P_l(\cos\theta)\exp\left[-\frac{l(l+1)Dt}{R^2}\right]\sin\theta d\theta \\
  =& R^2 \int_{-1}^1 (P_0(x) - P_1(x))\sum_{l=0}^\infty \exp\left[-\frac{l(l+1)Dt}{R^2}\right](2l+1)P_l(x) dx\\
  =& R^2 \sum_{l=0}^\infty \exp\left[-\frac{l(l+1)Dt}{R^2}\right](2l+1) \int_{-1}^1 (P_0(x) - P_1(x))P_l(x) dx \\
  =& R^2 \sum_{l=0}^\infty \exp\left[-\frac{l(l+1)Dt}{R^2}\right](2l+1) \left(2\delta_{l0} - \frac{2}{3}\delta_{l1} \right) \\
  =& 2R^2\left[ 1  - e^{-2Dt/R^2} \right]
\end{align*}
Note that in the limit for $t\rightarrow 0,~\left<\delta x^2\right> = 2R^2\left[1 - (1 - 2Dt/R^2) \right] = 4Dt,$ again the result for a 2D plane.

\section{Performance}
\label{sec:performance}
This section describes the performance of our implementation of {RATTLE} in LAMMPS. We compare how the algorithm performs against unconstrained velocity Verlet updates in a 2D at comparable densities. Additionally, we check its parallel efficiency on a small computing cluster which we again compare to the aforementioned unconstrained velocity Verlet scheme.

\begin{figure}[htb]
  \begin{center}
    \subcaptionbox{\label{fig:timing-vs-verlet}}{
      \includegraphics[width=0.47\textwidth]{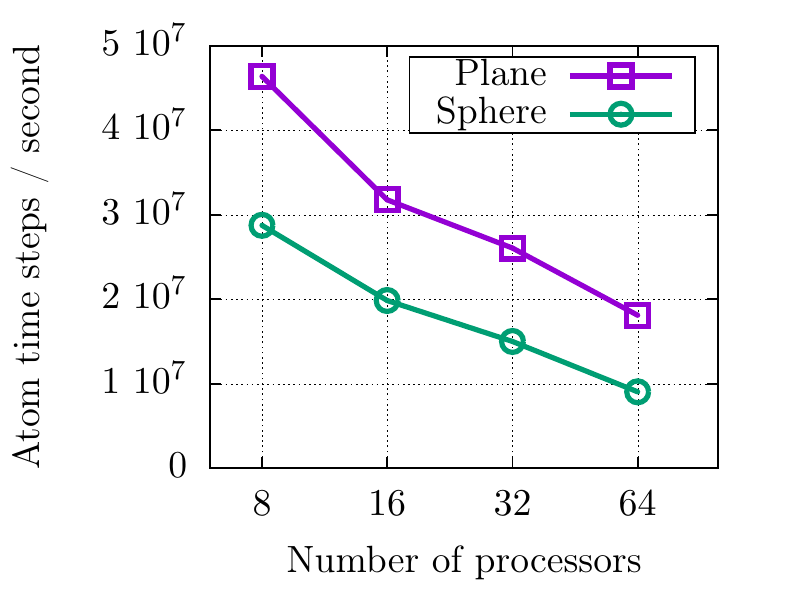}
    }
    \subcaptionbox{\label{fig:parr_eff}}{
      \includegraphics[width=0.47\textwidth]{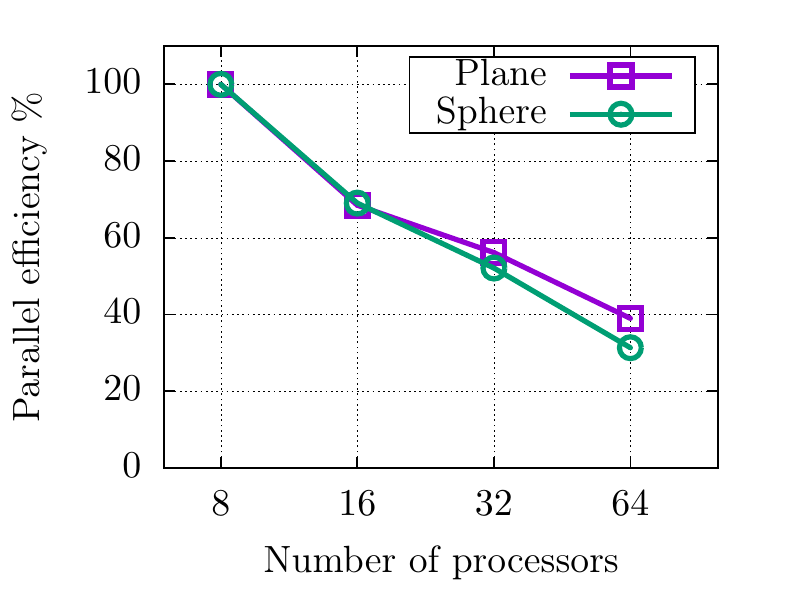}
    }
    \caption{Parallel performance (B) and parallel efficiency (B) of {RATTLE} applied to particles constrained to a sphere compared with an unconstrained, 2D velocity Verlet scheme. We time integrated 10000 particles per core in all cases. The parallel performance is expressed in atom time steps per second, which, in the ideal case, would be a constant, but is far from that on the compute cluster we used. The parallel efficiency is defined as the parallel performance for a given number of processors divided by the performance for 8 processors (one node). This quantity shows that {RATTLE} has a parallel scaling comparable to an unconstrained velocity Verlet algorithm, making it very suitable for simulating large systems. The difference for larger numbers of processors is because {RATTLE} requires slightly more communication between nodes.  \label{fig:parallel-scaling}}
  \end{center}
\end{figure}

To test the performance of {RATTLE}, we simulate a system of $N$ Lennard-Jones particles on a sphere of radius $R$ and on a 2D plane of size $2L$ by $2L$ with periodic boundary conditions. We tune the size of the system so that the densities of the sphere and plane match:
\begin{equation*}
  \phi = \frac{N}{4\pi R^2} = \frac{N}{4L^2}.
\end{equation*}
We vary the number of particles with the number of processors used in an attempt to keep the number of particles per processor constant to $N=10000$/core. This way information about the parallel efficiency is obtained as well, which we present later. For now, however, we only compare the computational cost of {RATTLE} with the standard velocity Verlet algorithm, both of which are plotted in figure \ref{fig:timing-vs-verlet}.
From the figure it is immediately clear that {RATTLE} loses time in performing the iterative scheme. The relative slowdown scales from 1.5 (meaning that {RATTLE} performs two time integration steps in the same time a velocity Verlet scheme performs 3) on single nodes up to 2 for simulations split across different nodes. Most of this slowdown, however, comes from poor communication performance of the small cluster we used for our benchmark, which will become apparent from in figure \ref{fig:parr_eff}.

Some attempts to optimise {our implementation of RATTLE} were considered. We tried different strategies to solve the Newton iteration scheme, and it was the fastest to solve the system by calculating the analytic solution to $\bvec{J}^{-1}\bvec{\Delta x} = \bvec{R}$ directly. Furthermore, a small speedup can be obtained by not updating the normal vector $\bvec{n}$ after each iteration. This has no noticble effect on energy conservation.

To gain more insight in the parallel scaling of {RATTLE}, we determine the so-called parallel efficiency. This measures how much slower a simulation twice as big spread over twice as many nodes is, compared to a reference system. In our case we studied the scaling from one to eight nodes. The results are shown in figure \ref{fig:parr_eff}, where we plot the parallel efficiency of {RATTLE} for a sphere and for the aforementioned unconstrained simulations. This reveals that the poor parallel scaling observed in figure \ref{fig:timing-vs-verlet} was actually due to the poor communication performance of the computing cluster used, rather than the result of an inefficient algorithm.

\section{Simulation setups}
This section contains the details relating to the examples presented in sections 3.2 and 3.3 in the article. 

\subsection{Crowded diffusion on curved membranes}
Here we describe the simulation details relating to the crowded diffusion example (Section 3.2). These simulations consisted of a varying number of one tracer particle and $N=2000,~2500,~3000,~3500$ and $4000$ crowder particles. The crowder particles and the tracer particle all interacted with the same interaction potential, namely a truncated shifted Lennard-Jones potential, obtained by putting $r_c = 2^{1/6}\sigma$ in equation (1). This steric repulsion models an excluded volume for each particle, of which the effect becomes more pronounced at higher particle numbers.

\paragraph{Constraint functions}

The surface they were constrained to consists of piece-wise continuous constraint functions of different shapes, namely planes, cylinder parts, and some ``connectors'' to make it possible for particles to smoothly diffuse from the blocks to the bridge. The constraint functions are given by
\begin{align*}
  g_\text{plane}(x,y,z)     =& a(x-x_0) + b(y-y_0) + c(z-z_0) \\
  g_\text{cylinder}(x,y,z)  =& a (x-x_0)^2 + b(y-y_0)^2 + c(z-z_0)^2 - R^2 \\
  g_\text{connector}(x,y,z) =& (x-x_0)^2 + \left(\sqrt{y^2+z^2} - (R_0+R)\right)^2 - R^2
\end{align*}
where $x_0,~y_0,~z_0,~a,~b,~c,~R,$ and $R_0$ are parameters that are different for each part. For the cylinders it is required that one of $a, b$ or $c=0$ while the other two are 1. The functional form of $g_\text{connector}$ may appear uninformative, but it is just the shortest distance of a point $(x,y,z)$ to a circle of radius $R$ in the $x,r(y,z)=\sqrt{y^2+z^2}$-plane with its centre at $(x=x_0,r(y,z) = R_0 + R).$ For more clarity, see the illustrations of the effectively obtained surface in figures 1a and 1b. The correct constraint function for each particle is selected based on its position with some simple programming logic. Finally, we present here the normal vectors for each constraint function as well:
\begin{align*}
  \bvec{n}_\text{plane}(x,y,z)     =& a\bvec{e}_x + b\bvec{e}_y + c\bvec{e}_z \\
  \bvec{n}_\text{cylinder}(x,y,z)  =& 2a (x-x_0)\bvec{e}_x + 2b(y-y_0)\bvec{e}_y + 2c(z-z_0)\bvec{e}_z \\
  \bvec{n}_\text{connector}(x,y,z) =& 2(x-x_0) \bvec{e}_x + 2\left(1 - \frac{R_0 + R}{\sqrt{y^2+z^2}}\right)(y\bvec{e}_y + z\bvec{e}_z)
\end{align*}
with $\bvec{e}_{x,y,z}$ unit vectors pointing in the $x,$ $y$ and $z$ directions.
For the bridge, we used a cylinder with either $R = R_b = 1.5\sigma$ or $2.5\sigma.$ The connectors were made to match this by putting $R_0 = R_b,$ while $R=3\sigma$ for both values of $R_b.$ The length of the cylindrical part of the bridge was $10\sigma,$ so in combination with the connectors the total distance between the two blocks is $16\sigma.$ The radii of the cylindrical parts in the blocks were also $3\sigma.$ For the cylindrical part of the bridge, $a=0$ and $b=c=1,$ while for those in the blocks $a=b=1$ and $c=0.$ For the planes we determine $x_0,~y_0$ and $z_0$ based on where they join with a cylinder, and we choose the signs of $a~b$ and $c$ so that the sign of $\bvec{n}$ is equal for the plane and the joining cylinders and connectors.

\paragraph{Surface area}

To determine the area coverage $\phi$ one needs to know both an effective area for the particles and the total area of the two blocks and the connecting bridge, say $A_t.$ The blocks consist of four quarter cylinders of equal radius, say $R_c,$ and four planes of equal size, say $L_x \times L_z.$ Let the cylinder and plane areas be $A_c$ and $A_p,$ respectively. The bridge is another cylinder with a different radius, say $R_b,$ and a length $L_b$ and has an area $A_b.$ Finally, there are two connectors between the cylinder and the blocks, for which the area $A_{conn}$ can be determined with an integral. However, this connector also blocks an area of $\pi (R_0 + R)^2$ of the blocks, for which we have to correct. The total area is thus $A_t = 2A_c + 8A_p - 2 A_{conn} + A_b.$ The $z$-dimension of the simulation volume extended over a length $L_z=30\sigma,$ and the volume was periodic in this dimension. Hence, the areas of the cylinders and planes in the blocks are $A_c = 2\pi L_z R_c$ and $A_b = L_z L_x.$ The bridge has an area of $A_b = 2 \pi L_b R_b.$ The connector is a curve above the x-axis given by $r_{y,z}(x):= \sqrt{y^2 + z^2} = R_b + R\left[1 - \sqrt{1 - ((x-x_0)/R)^2} \right].$ An infinitesimal area element of this curve revolved around the x-axis is thus given by
\begin{equation*}
  dA_{conn} = r_{y,z}(x) \sqrt{ 1 + \left( \frac{\partial r_{y,z}}{\partial x}\right)^2 } dx d\phi,
\end{equation*}
The derivative is easily determined to be
\begin{equation*}
  \frac{\partial r_{y,z}}{\partial x} = \frac{1}{\sqrt{1 - ((x-x_0)/R)^2}} (x-x_0)/R = \frac{(x-x_0)}{\sqrt{R^2 - (x-x_0)^2}}
\end{equation*}
and from this one can show that $1 + (\partial r_{y,z}/\partial x)^2 = 1/[1 - (x-x_0)^2/R^2].$
Combining all terms and a substitution of $u = (x-x_0)/R$ leads to the following total area:
\begin{align*}
  A_{conn} =& \int_{\phi=0}^{2\pi}d\phi\int_{u=0}^{1} \frac{R_b + R\left(1 - \sqrt{1 - u^2}\right)}{\sqrt{1 - u^2}}R du \\
  =& 2\pi \int_{u=0}^1 \left[R \frac{R_b + R}{\sqrt{1-u^2}} - R^2 \right]du = 2 \pi R(R+R_b) \frac{\pi}{2} - 2 \pi R^2 \\
  =& 2\pi R \left[ \pi\frac{R + R_b}{2} - R \right] = \pi R^2 \left( \pi - 2  \right) + \pi^2 R R_b
\end{align*}
Thus, in conclusion, the total area of the entire curved surface is given by $A_t = 2A_c + 8A_p - 2 A_{conn} + A_b,$ with
\begin{align*}
  A_c =& 2\pi R_c L_z, \qquad A_p = L_x L_z, \qquad A_b = 2 \pi L_b R_b, \\
  A_{conn} =&\pi R^2 \left( \pi - 2 \right) + \pi^2 R R_b.
\end{align*}
We generated data for both $R_b = 1.5 \sigma$ and $R_b = 2.5 \sigma,$ with $\sigma$ the characteristic Lennard-Jones distance. We kept the other parameters constant to the values listed in table \ref{tab:params}

\begin{table}[hbt]
  \centering
  \caption{Surface parameters used in the crowded diffusion on curved membranes setup. All units are expressed per particle diameter $\sigma.$ \label{tab:params}}
  \begin{tabular}{ccccc}
    $L_x$ & $L_z$ & $R_c$ & $L_b$ & $R$ \\ \hline
    15 & 30 & 3 & 10 & 3
  \end{tabular}
\end{table}

\subsection{Virus capsid self-assembly}
Here we describe the simulation details relating to the virus capsid self-assembly example (Section 3.3).
Because we constrain only one bead in the conical particle, it is in principle possible that they flip ``outside-in''. To prevent this, we gently push beads out of the sphere centre with a repulsive Lennard-Jones wall, whose potential is
\begin{align*}
  V_{w} =4\epsilon_w \left[ \left( \frac{\sigma_w}{r - R_w} \right)^{12} - \left( \frac{\sigma_w}{r - R_w} \right)^6 + \frac{1}{4} \right] \cdot H( 2^{1/6}\sigma_w - ( r - R_w ) ),
\end{align*}
with $H(x)$ the Heaviside function which is 1 if $x>0$ and 0 otherwise.
The adjacent beads in a cone are bonded with a harmonic potential $V_\text{bond},$ and we invoke an angular potential $V_\text{bend}$ for three adjacent beads:
\begin{align*}
  V_\text{bond}(\bvec{x}_i, \bvec{x}_j) =& \kappa_{b} (\left\|\bvec{x}_i - \bvec{x}_j\right\| - r_{0,ij})^2, \quad
  V_\text{bend}(\bvec{x}_i, \bvec{x}_j, \bvec{x}_k) = \kappa_{a} (\theta - \theta_0)^2, \\
  \theta =& \arccos\left[ (\bvec{x}_i - \bvec{x}_j)\cdot (\bvec{x}_j - \bvec{x}_k) / \left(\left\|\bvec{x}_i - \bvec{x}_j\right\| \left\|\bvec{x}_j - \bvec{x}_k \right\|\right) \right]
\end{align*}
The coefficients $r_{0,ij}$ depend on the bead types $i$ and $j$ their sizes $\sigma_i,\sigma_j$ as $r_{0,ij} = 2^{1/6}(\sigma_i + \sigma_j)/4.$ $\kappa_b$ was constant at $50k_BT/\sigma_0^2$ for all bonds, with $\sigma_0$ the size of the smallest bead. Furthermore, $\theta_0$ was $\pi$ for both angles and $\kappa_a=250k_BT/ \mathrm{rad}^2.$ The bead sizes are given in table \ref{tab:coeffs}. The well depth was $4k_B T$ in all cases. We use an additive mixing rule: $\sigma_{ij} = \half(\sigma_i + \sigma_j).$ Masses scaled according to volume, with the mass of the smallest bead 1. For all  interactions except those between like beads of types 2 and 3 were purely repulsive ($r_c = 2^{1/6} \sigma_{ij}$). For types $i=2,~3,$ $r_c = 4\sigma_{0}.$ The Lennard-Jones interactions are only applied between beads in different cones.

\begin{table*}[b]
  \centering
  \caption{Table of coefficients dependent on bead size, all in units of $\sigma_0$}
  \label{tab:coeffs}
  \begin{tabular}{cccc}
    $\sigma_1$ & $\sigma_2$ & $\sigma_3$ & $\sigma_4$  \\ \hline
    1 & 1.35 & $[1.4 - 2.1]$ & 1
  \end{tabular}
\end{table*}

\end{document}